\documentclass{PoS}

\usepackage{graphicx}
\usepackage{amsmath}
\usepackage{amssymb}
\usepackage{multirow}

\DeclareGraphicsExtensions{.pdf,png,.jpg}

\newcommand{\apjl}{{Astrophys. J. Lett.} }

\newcommand{\app}{{Astropart. Phys. }}

\newcommand{\etal}{et al., }
\newcommand{\nima}{{Nucl. Instrum. Methods Phys. Res., Sect. A }}

\title{Future Ground-based Wide Field of View Air Shower Detectors}

\ShortTitle{Wide Field of View Detectors}

\author{\speaker{Di Sciascio Giuseppe}\\
        INFN - Roma Tor Vergata\\
        E-mail: \email{disciascio@roma2.infn.it}}

\abstract{ 

Extensive air shower (EAS) arrays directly sample the shower particles that reach the observation altitude. 
They are wide field of view (FoV) detectors able to view the whole sky simultaneously and continuously. In fact, EAS arrays have an effective FoV of about 2 sr and operate with a duty cycle of $\sim$100\%.
This capability makes them well suited to study extended sources, such as the Galactic diffuse emission and measure the spectra of Galactic sources at the highest energies (near or beyond 100 TeV).
Their sensitivity in the sub-TeV/TeV energy domain cannot compete with that of Cherenkov telescopes, but the wide FoV 
is ideal to perform unbiased sky surveys, discover transients or explosive events (GRBs) and monitor variable or flaring sources such as Active Galactic Nuclei (AGN). 
An EAS array is able to detect at the same time events induced by photons and charged cosmic rays, thus studying the connection between these two messengers of the non-thermal Universe. 
Therefore, these detectors are, by definition, multi-messenger instruments. 

Wide FoV telescopes are crucial for a multi-messenger study of the Gravitational Wave events due to their capability to survey simultaneously all the large sky regions identified by LIGO and VIRGO, looking for a possible correlated $\gamma$-ray emission.

In this contribution we summarize the scientific motivations which push the construction of new wide FoV air shower detectors and introduce the future instruments currently under installation. Finally, we emphasize the need of an EAS array in the Southern hemisphere to monitor the Inner Galaxy and face a number of important open problems.
}

\FullConference{XII Multifrequency Behaviour of High Energy Cosmic Sources Workshop\\
		12-17 June, 2017\\
		Palermo, Italy}

\begin{document}

\section{Open problems in Cosmic Ray Physics}

Understanding the origin of the \emph{"knee"} in the energy spectrum of the primary radiation is the key for a comprehensive theory of the origin of Cosmic Rays (hereafter CR) up to the highest observed energies.
In fact, the knee is clearly connected with the issue of the end of the Galactic CR spectrum and the transition from Galactic to extra-galactic CRs.

If the knee, a steepening of the spectral index from $\sim$-2.7 to $\sim$-3.1 at about 3 PeV (=3$\times$10$^{15}$ eV), is a source property we should see a corresponding spectral feature in the gamma-ray spectra of the CR sources. If, on the contrary, this feature is the result of propagation, we should observe a knee that is potentially dependent on location, because the propagation properties depend, in principle, on the position in the Galaxy.

To understand the origin of the knee we need to deepen our understanding of acceleration, escape and propagation of the relativistic particles, the main pillars that constitute the SuperNova paradigm for the origin of the radiation (see \cite{morlino17} and references therein).
We need to identify the sources and the mechanisms able to accelerate particles beyond PeV energies (the so-called \emph{"PeVatrons"}). We need to understand how particles escape from the sources and are released into the interstellar medium. Finally, we need to understand how particles propagate through the Galaxy before reaching the Earth.

It is widely believed that the bulk of CRs up to about 10$^{17}$ eV are Galactic, produced and accelerated by the shock waves of SuperNova Remnants (SNR) expanding shells \cite{drury12}.
The SNR paradigm has two bases: firstly, the energy released in SN explosions can explain the CR energy density considering an overall efficiency of conversion of explosion energy into CR particles of the order of 10\% . 
Secondly, the diffusive shock acceleration operating in SNR can provide the necessary power-law spectral shape of accelerated particles with spectral index -2.0 that subsequently steepen to -2.7, as observed, due to the energy-dependent diffusive propagation effect (see \cite{drury17} and references therein).

SuperNovae are believed to be almost the only available power source. 
However, recent claims by H.E.S.S. of a possible detection of a PeVatrons in the Galactic Center, most likely related to a supermassive black hole \cite{hess-pev}, opens new perspectives showing that galactic PeVatrons other than SNRs may exist.

Recently AGILE and Fermi observed GeV photons from two young SNRs (W44 and IC443) showing the typical spectrum feature around 1 GeV (the so-called \emph{'$\pi^0$ bump'}, due to the decay of $\pi^0\to\gamma\gamma$) related to hadronic interactions \cite{pizero-a,pizero-f}. 
This important measurement, however, does not demonstrate the capability of SNRs to produce the power needed to maintain the galactic CR population and to accelerate CRs up to the knee, at least. 
In fact, unlike neutrinos that are produced only in hadronic interactions, the question whether $\gamma$-rays are produced by the decay of $\pi^0$ from protons or nuclei interactions (\emph{'hadronic'} mechanism), or by a population of relativistic electrons via Inverse Compton scattering or bremsstrahlung (\emph{'leptonic'} mechanism), still needs a conclusive answer.

One of the main open problems in the SNR origin model is the maximum energy that can be attained by a CR particle in SNR. 
To accelerate protons up to the PeV energy domain a significant amplification of the magnetic field at the shock is required but this process is problematic in SNRs \cite{gabici16}. 
However, if the knee is a propagation effect, the Galaxy could contain "super-PeVatrons", sources capable to accelerate particles well beyond the PeV. The study of these objects requires to observe the $\gamma$-ray sky at 100 TeV energies and beyond.

No direct observational evidence for the acceleration of PeV protons in SNRs has been reported yet, probably due to the fact that higher energy (> 200 TeV) particles are believed to be accelerated in the early phases of the SuperNova explosion (i.e. in young SNRs). Therefore, we expect that very few SNRs are currently accelerating particles up to PeV energies.
In addition, the absorption of $\gamma$-rays may prevent observations of PeVatrons \cite{vernetto17}.
Finally, the sensitivity of current gamma-ray detectors above 100 TeV is very poor. 

As we will discuss in the next Sections, the expected sensitivity of next generation wide FoV detectors like LHAASO and HiSCORE will open for the first time the PeV range to observations.

Understanding the CR origin and propagation at any energy is made difficult by the poor knowledge of the elemental composition of the radiation as a function of the energy.
In the next future ISS-CREAM on the ISS \cite{isscream} will approach the PeV range, but to clarify the nature of the knee an integrated measurement of the evolution of the heavy component, and of the CR anisotropy separately for different primary masses, across the knee region is mandatory.

An integrated and statistically significant measurement of the energy spectrum, elemental composition and anisotropy in the PeV energy region can be carried out only by ground-based EAS arrays.
In fact, since the CR flux rapidly decreases with increasing energy and the size of detectors is constrained by the weight that can be placed on satellites/balloons, their collecting area is small and determines a maximum energy (of the order of a few hundred TeV/nucleon) related to a statistically significant detection. In addition, the limited volume of the detectors makes difficult the containement of showers induced by high energy nuclei, thus limiting the energy resolution of instruments in direct measurements.

Working at high altitude ($>$4000 m asl) is crucial to study CRs in the knee region because: (1) the shower fluctuations are reduced since the detector approaches the depth of the maximum longitudinal development; (2) the average shower size in the knee energy region is almost the same for all nuclei, providing a composition independent estimator of the energy; (3) the reduced attenuation in the atmospheric overburden permits the observation of EAS induced by primaries with TeV energies, thus allowing the calibration of the absolute energy scale exploiting the \emph{"Moon shadow"} technique, as demonstrated by the ARGO-YBJ experiment  \cite{argo-moon}. In addition the superposition with direct measurements allows a cross-calibration of the different detectors. The calibration of the relation between shower size and primary energy is one of the most important problem for ground-based measurement, heavily affecting the reconstruction of the CR energy spectrum.

In the standard picture, mainly based on the results of the KASCADE esperiment, the knee is attributed to the steepening of the p and He spectra \cite{kascade}. 
According to a rigidity-dependent structure (Peters cycle), the sum of the fluxes of all elements, with their individual knees at energies proportional to the nuclear charge E$_Z$ = Z$\times$ 3 PeV, makes up the CR all-particle spectrum \cite{peters}.
With increasing energies not only the spectrum becomes steeper, due to such cutoffs, but also heavier.
%
\begin{figure}
\centerline{\includegraphics[width=0.8\textwidth,clip]{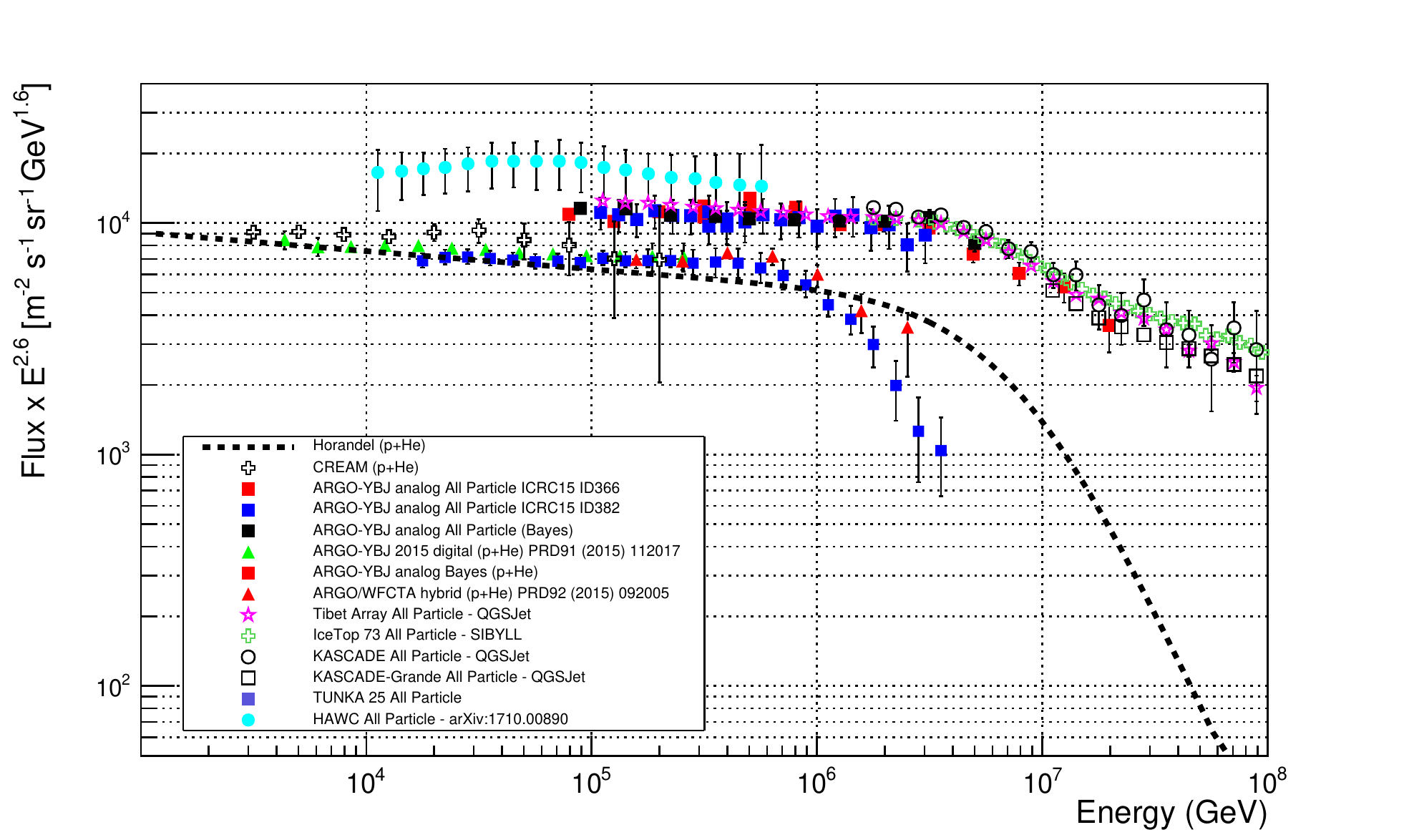} }
\caption{All-particle and light (p+He) component energy spectra of primary CR measured by ARGO-YBJ and compared to different experimental results. The parametrization provided by Horandel \cite{horandel} is shown for comparison. The systematic uncertainty is shown by the error bars.}
\label{fig:argo-phe-knee}       
\end{figure}
%
However, a number of results (in particular those obtained by experiments located at high altitudes) seem to indicate that the bending of the light component (p+He) is well below the PeV and the knee of the all-particle spectrum is due to heavier nuclei \cite{tibet,casamia,basje-mas}. 

Recent results obtained by the ARGO-YBJ experiment (located at 4300 m asl) clearly show, with different analyses, that the knee of the light component starts at $\sim$700 TeV, well below the knee of the all-particle spectrum that is confirmed by ARGO-YBJ at $\sim$4$\cdot$10$^{15}$ eV \cite{hybrid15} (see Fig. \ref{fig:argo-phe-knee}).

After more than half a century from the discovery of the knee experimental results are still conflicting with uncertainties on its origin.
This is not surprising for a lot of reasons.
The reconstruction of the CR elemental composition is often carried out by means of complex unfolding techniques based on the measurements of electronic and muonic sizes, procedures that heavily depend on the hadronic interaction models. 
The muonic size is much smaller than the electronic one with a wider lateral distribution, but the total sensitive area of muon detectors is typically only few hundred square meters and, due to the poor sampling, large instrumental fluctuations can be added to the stochastic ones associated to the shower development. In addition, the \emph{'punch-through effect'} due to high energy secondary electromagnetic particles could heavily affect the measurements.
Finally, some arrays have been operated close to the sea level and not in the shower maximum region.

With the full coverage array operated in the ARGO-YBJ experiment at the 4300 m asl, very close to the shower maximum, for the first time the shower core region has been studied with unprecedented details \cite{argo-core}. In addition, events are selected on an event-by-event basis exploiting the lateral distribution of secondary particles and not the muonic size, with results nearly independent from the hadronic interaction models.

The observation of a Helium energy spectrum harder than the proton one, reported by Pamela \cite{pamelahe}, CREAM \cite{creamhe} and AMS02 \cite{ams02he}, has the interesting consequence that the knee region could be dominated by Helium and CNO masses with a proton knee below the PeV, as suggested by the ARGO-YBJ results.

The measurement of the anisotropy in the arrival direction distribution of CRs is a complementary way to understand the origin and the propagation of the radiation. In fact CR anisotropy is a fingerprint for their origin and propagation \cite{disciascio13,disciascio15,ahlers16}.
The study of the anisotropy can clarify the origin of the knee. Indeed, if the knee is due to an increasing inefficiency in CR containment in the Galaxy a change in anisotropy is expected. If, on the contrary, the knee is related to the limit of the acceleration mechanism, we do not expect a change of anisotropy across the knee.

\begin{figure}[t!]
  \centerline{\includegraphics[width=0.8\textwidth]{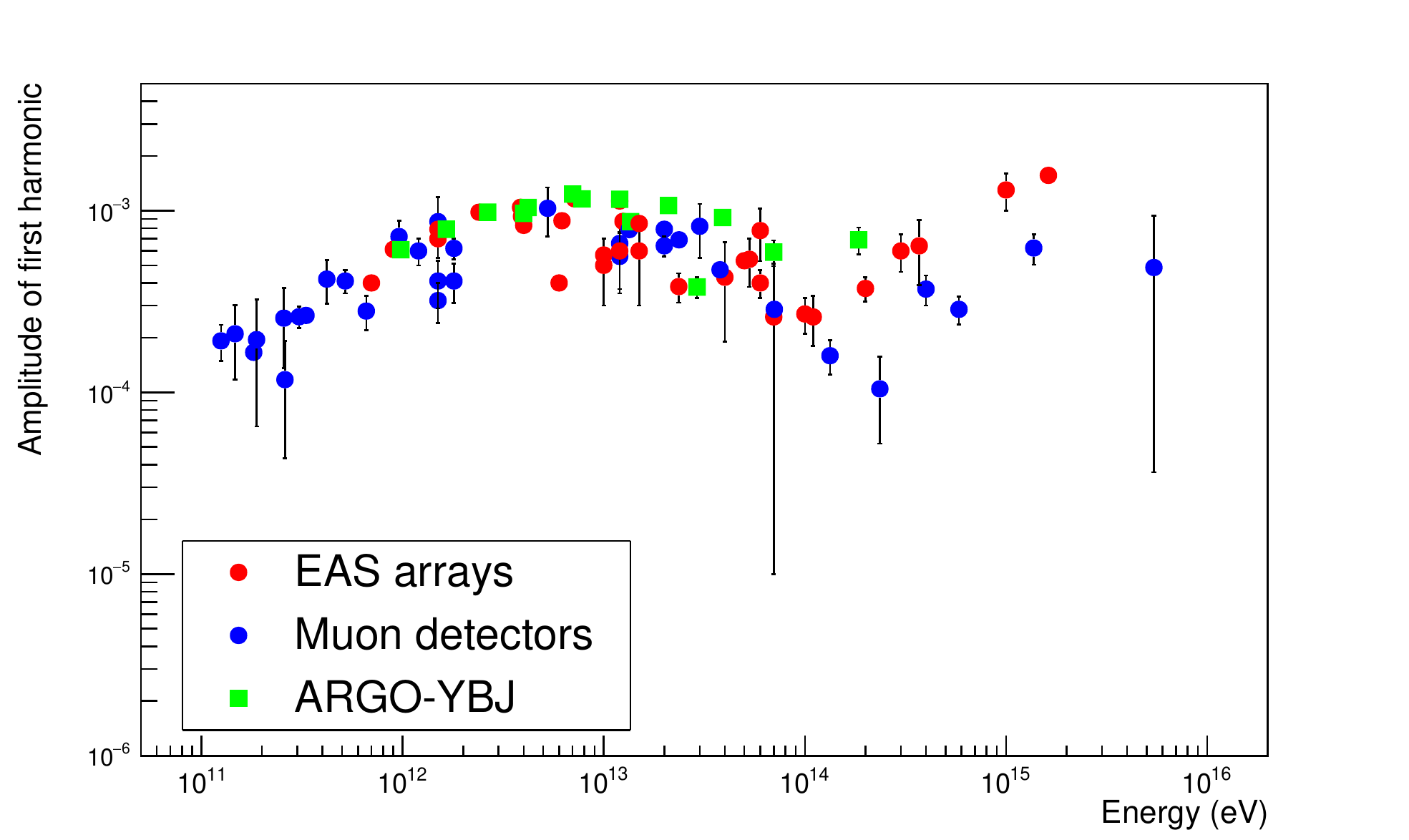}}
    \centerline{\includegraphics[width=0.8\textwidth]{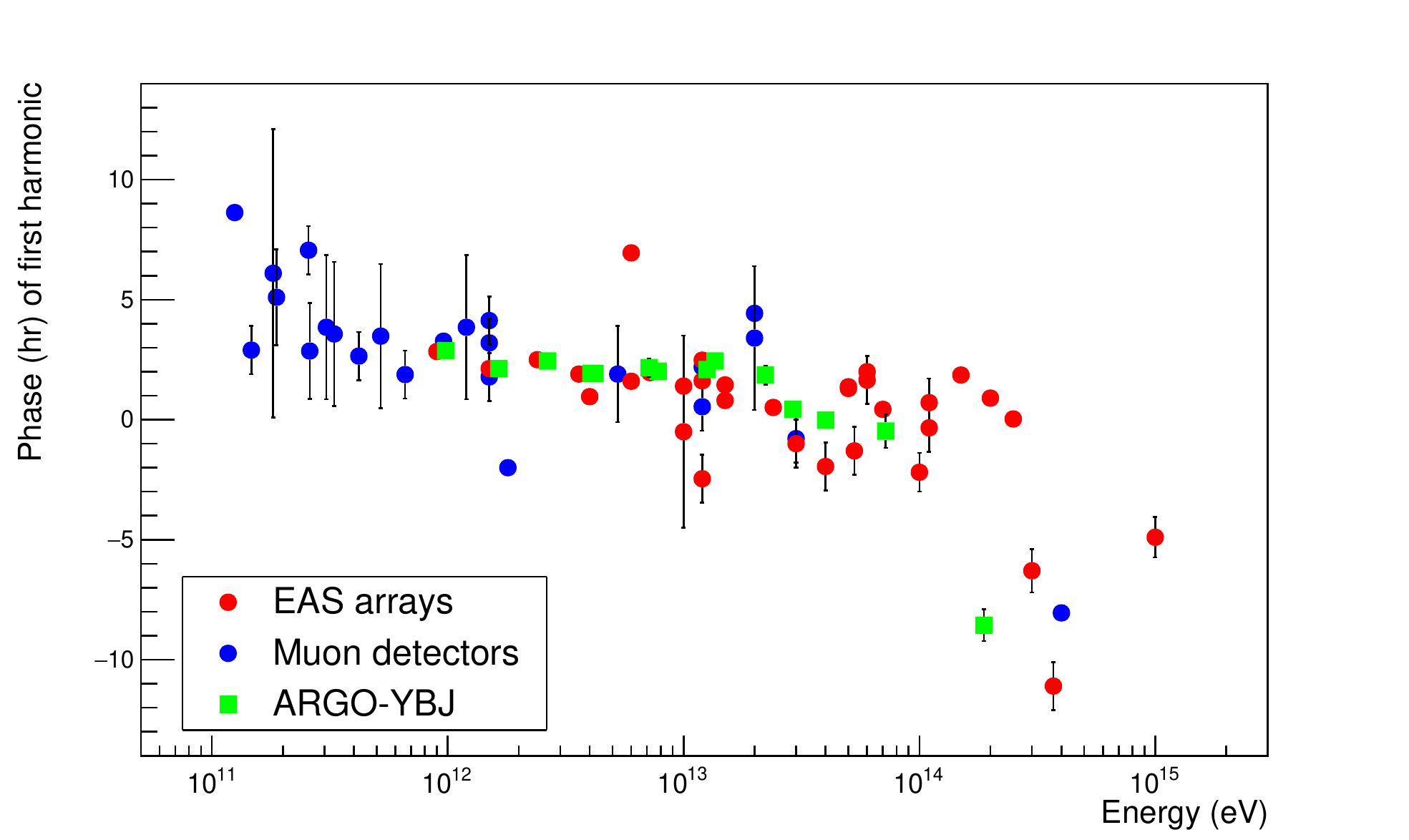}}
\caption[h]{Amplitude and phase of the first harmonic (upper and lower plots, respectively) measured by ARGO-YBJ \cite{argo-lsa,argo-lsaicrc17} compared with a compilation of data obtained by different experiments (muon detectors or EAS-arrays) as a function of the CR primary energy (for details and references see \cite{disciascio13}).} 
\label{fig:amplit-phase}
\end{figure}

The unexpected variation of the anisotropy with energy (Fig. \ref{fig:amplit-phase}) seems to be difficult to interpret in terms of conventional Galactic CR diffusion in the Galaxy.
%
%
The dramatic change of the anisotropy phenomenology above 100 TeV may provide important information for understanding the origin and propagation of the particles.
A statistically significant measurement of the evolution of anisotropy across the knee region for different primary masses is crucial to disentangle between different models of CR propagation in the Galaxy \cite{argo-anismass}.

\section{Motivation for a wide field of view $\gamma$-ray telescope}
\label{ga}

Charged cosmic rays, photons and neutrinos are strongly correlated in CR sources where hadronic accelerators are at work. 
Their integrated study is one of the most important and exciting fields in the \emph{'multi-messenger astronomy'}, the exploration of the Universe through combining information from different cosmic messengers: electromagnetic radiation, gravitational waves, neutrinos and cosmic rays.

In a hadronic interaction the secondary photons have an energy a factor of 10 lower than the primary proton. 
Therefore, the quest for CR sources able to accelerate particles in the knee energy region requires to survey the $\gamma$-ray sky above 100 TeV. In addition, the Inverse Compton scattering at these energies is strongly suppressed by the Klein-Nishina effect. Therefore, the observation of a $\gamma$-ray power law spectrum extending up to the 100 TeV range would be a strong indication of the hadronic nature of the emission.

So far no photons above 100 TeV have been observed from any source, and only six sources have data above 30 TeV: the SNR RX J1713.7-3946, the Pulsar Wind Nebula (PWN) Crab and Vela-X, and the extended sources MGROJ2031+41, MGROJ2019+37 and MGROJ1908+06, all of them probably PWN too.
Their spectra above 30 TeV is, however, only known with large uncertainties, being the sensitivity of the current instruments at the highest energies not enough to determine clearly the spectral shape. 

A problem that cannot be neglected when working at high energy, is the absorption of gamma rays due to pair production in the interstellar and intergalactic space.  High energy $\gamma$-rays interact with the infrared/optical photons and with the Cosmic Microwave Background radiation (CMB).
The absorption increases with the $\gamma$-ray energy and the source distance, being particularly effective for extragalactic sources,  but at sufficiently high energy can also affect the flux of galactic objects. 
Concerning galactic sources, the absorption mainly depends on the relative position of the source and the Sun inside the Galaxy.
Up to $\sim$20 TeV the flux attenuation is less than 3\% for any position. Above $\sim$20--30 TeV the absorption increases due to the interaction with infrared photons, with a maximum effect at $\sim$150 TeV. At this energy $\sim$30--50\% of photons from sources located beyond the Galactic center, are absorbed. Above $\sim$200 TeV the CMB becomes the major cause of absorption, whose amount only depends on the source distance. At these energies, the absorption can be a severe obstacle to observations of sources at distances larger than a few kiloparsecs \cite{vernetto17,vernetto-icrc17}.

To open the 100 TeV range to observations a detector with a very large effective area, operating with high duty-cycle, is required.
The most sensitive experimental technique for the observation of $\gamma$-rays at these energies and above is the detection of EAS via large ground-based arrays.
The muon content of photon-induced showers is very low, therefore these events can be discriminated from the large background of CRs via a simultaneous detection of muons that originate in the muon-rich CR showers. 
In addition, the large FoV and high duty cycle of EAS arrays make this observational technique particularly suited to perform unbiased all-sky surveys, and not simply observations of limited regions of the Galactic plane.

EAS arrays are also extremely valid instruments in the sub-TeV/TeV energy domains.
Their sensitivity in this energy range cannot compete with that of the future generation Cherenkov telescopes, as CTA (that also have a better energy and angular resolution) \cite{cta-science}, but the wide FoV offers the opportunity to continuously monitor a large fraction of the sky, looking for unpredictable transient phenomena. 
In fact, most current Cherenkov telescopes can work only during clear moonless nights, with a total observation time of about 1000--1500 hours per year (depending on the location), with a typical duty cycle of $\sim$10--15$\%$, and a FoV of a few degrees radius. This implies that they can observe only one source at the time, and only in the season of the year when the source culminates during night time.

By contrast, an EAS detector every day can observe a large fraction of the celestial sphere (spanning 360$^{\circ}$ in right ascension and about 90$^{\circ}$ in declination, in a declination interval depending on the geographic location). 
Sources located in this portion of the sky are in the FoV of the detector, either always, or for several hours per day, depending on their celestial declination.
This situation is ideal to perform sky surveys, discover transients or explosive events, such as Gamma Ray Bursts, and monitor variable or flaring sources such as Active Galactic Nuclei.

Another important topic that can be successfully addressed by a wide FoV detector is the extended $\gamma$-ray emission and in particular the study of the galactic diffuse gamma emission. This radiation, if produced by protons and nuclei via the decay of $\pi^0$ generated in hadronic interactions with interstellar gas, can trace the location of the CR sources and the distribution of interstellar gas.
In addition, the observation of a knee in the energy spectrum of the diffuse emission, at an energy of a few hundred TeV, corresponding to the knee of the CR all-particle spectrum, would provide a complementary way to investigate the origin of the knee \cite{vernetto-icrc17}. 
A new wide FoV experiment should also be able to map the diffuse emission in different regions of the Galactic Plane to observe a location dependence of the $\gamma$-ray spectral index \cite{argo-diffuse} and/or of the knee energy.

Wide FoV telescopes are crucial for a multi-messenger study of the Gravitational Wave events due to their capability to survey simultaneously all the large sky regions identified by LIGO and VIRGO, looking for a possible correlated $\gamma$-ray emission.
As an example, the event GW170817 was localized within a 3 by 10 degree region, well beyond the typical FoV of a Cherenkov telescope and required multiple pointings to H.E.S.S. to cover the area \cite{hess-gw}.

\section{Future Wide Field of View Detectors}

The only wide FoV detector currently in data taking to study gamma-ray astronomy in the sub-TeV/multi-TeV energy domain is HAWC. 

The HAWC (High Altitude Water Cherenkov) detector is located on the Sierra Negra volcano in central Mexico at an elevation of 4100 m a.s.l. 
It consists of an array of 300 water Cherenkov detectors made from 5 m high, 7.32 m diameter, water storage tanks  covering an instrumented area of about 22,000 m$^2$ (the actual tank coverage is 12,550 m$^2$ with a coverage factor less than 60\%).
Four upward-facing photomultiplier tubes (PMTs) are mounted at the bottom of each tank: a 10'' PMT positioned at the center, and three 8'' PMTs positioned halfway between the tank center and rim. 
The central PMT has roughly twice the sensitivity of the outer PMTs, due to its superior quantum efficiency and larger size. The WCDs are filled to a depth of 4.5 m, with 4.0 m (more than 10 radiation lengths) of water above the PMTs. This large depth guarantees that the electromagnetic particles in the air shower are fully absorbed by the HAWC detector, well above the PMT level, so that the detector itself acts as an electromagnetic calorimeter providing an accurate measurement of e.m. energy deposition \cite{hawc12,hawc1,hawc2}.

After a couple of years of operations a number of interesting results has been obtained.
In this contribution we mention just the publication of the 2nd catalog of gamma-ray sources \cite{hawc-2ndcat} that demonstrates the enormous potential of wide FoV detectors in discovering new $\gamma$-ray sources.
Realized with 507 days of data, with an instantaneous FoV $>$1.5 sr and $>$90\% duty cycle, represents the most sensitive TeV survey to date for such a large fraction of the sky. 
The median energy of the lowest multiplicity bin is about 700 GeV. 
A total of 39 sources were detected, with an expected contamination of 0.5 due to background fluctuations. Out of these sources, 19 are new sources that are not associated with previously known TeV sources. Ten are reported in TeVCat as PWN or SNR: 2 as blazars and the remaining eight as unidentified.

Two new wide FoV detectors, LHAASO and TAIGA-HiSCORE, are under installation and will be briefly described in the following.

\subsection{The LHAASO experiment}
A new project, developed starting from the experience of the high altitude experiment ARGO-YBJ \cite{discia-rev},  is LHAASO \cite{lhaaso1}. 
The experiment is strategically built to study with unprecedented sensitivity the energy spectrum, the elemental composition and the anisotropy of CRs in the energy range between 10$^{12}$ and 10$^{17}$~eV, as well as to act simultaneously as a wide aperture ($\sim$2 sr), continuosly-operated gamma-ray telescope in the energy range between 10$^{11}$ and $10^{15}$~eV.
%
\begin{figure}
\centerline{\includegraphics[width=0.8\textwidth,clip]{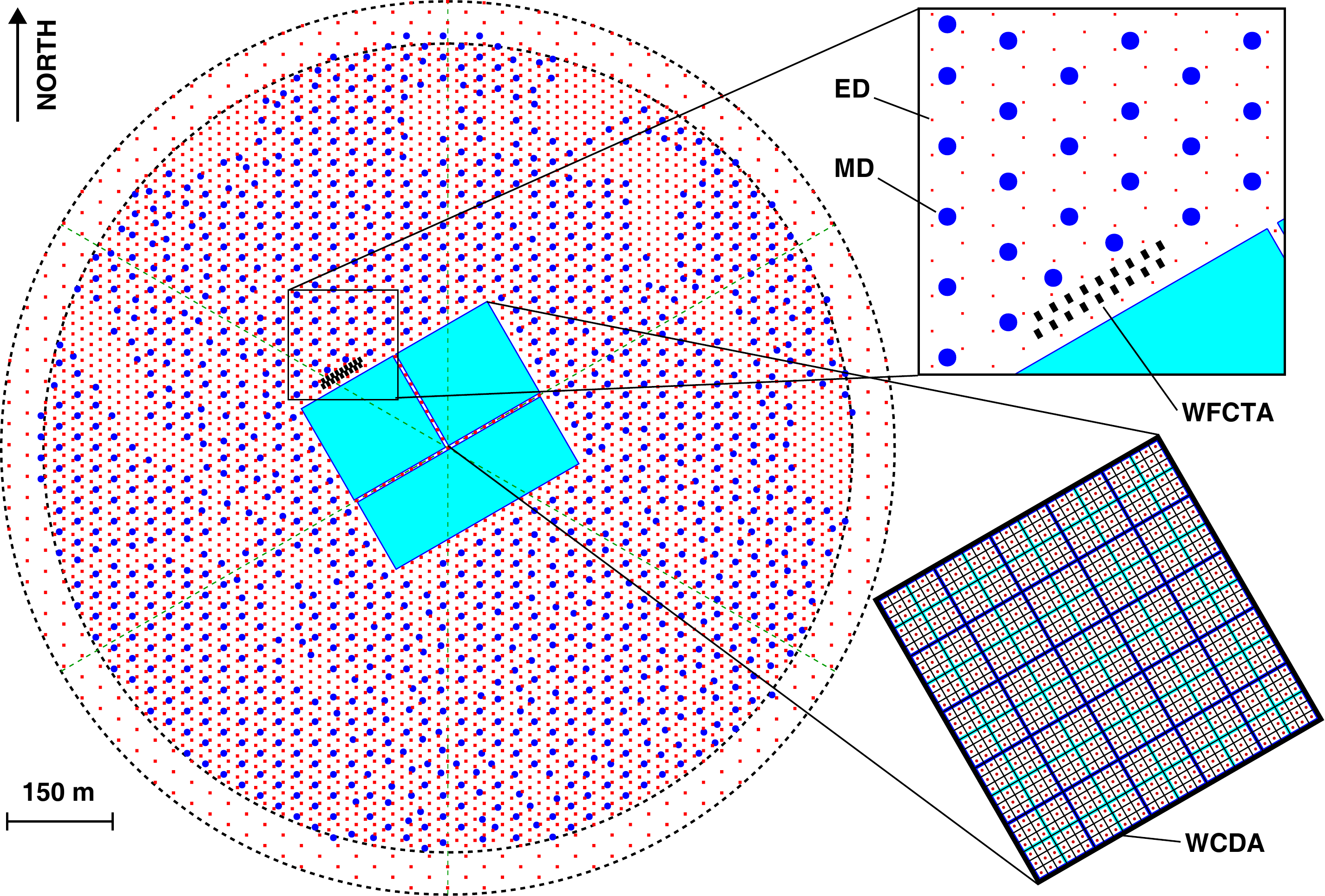} }
\caption{Layout of the LHAASO experiment. The insets show the details of one pond of the WCDA and of the KM2A array constituted by two overimposed arrays of electromagnetic particle detectors (ED) and of muon detectors (MD). The telescopes of the WFCTA, located at the edge of a pond, are also shown.} 
\label{fig:lhaaso-layout}       
\end{figure}
%

The first phase of LHAASO will consist of the following major components (see Fig. \ref{fig:lhaaso-layout}):
\begin{itemize}
\vspace{-0.2cm}
\item 1.3 km$^2$ array (LHAASO-KM2A) for electromagnetic particle detectors (ED) divided into two parts: a central part including 4931 scintillator detectors 1 m$^2$ each in size (15 m spacing) to cover a circular area with a radius of 575 m and an outer guard-ring instrumented with 311 EDs (30 m spacing) up to a radius of 635 m.
\vspace{-0.2cm}
\item An overlapping 1 km$^2$ array of 1146 underground water Cherenkov tanks 36 m$^2$ each in size, with 30 m spacing, for muon detection (MD, total sensitive area $\sim$42,000 m$^2$).
\vspace{-0.2cm}
\item A close-packed, surface water Cherenkov detector facility with a total area of about 78,000 m$^2$ (LHAASO-WCDA).
\vspace{-0.2cm}
\item 18 wide field-of-view air Cherenkov telescopes (LHAASO-WFCTA).
\end{itemize}

LHAASO is under installation at high altitude (4410 m asl, 600 g/cm$^2$, 29$^{\circ}$ 21' 31'' N, 100$^{\circ}$ 08'15'' E) in the Daochen site, Sichuan province, P.R. China. 
The commissioning of one fourth of the detector will be implemented in 2018 with an expected sensitivity similar to HAWC.
The completion of the installation is expected by the end of 2021.

In Table 1 the characteristics of the LHAASO-KM2A array are compared with other experiments. As can be seen, LHAASO will operate with a coverage of $\sim$0.5\% over a 1 km$^2$ area.
The sensitive area of muon detectors is unprecedented and about 17 times larger than CASA-MIA, with a coverage of about 5\% over 1 km$^2$.
By using LHAASO-WCDA as a further muon detector, the total sensitive area for $\mu$-detection will be about 120,000 m$^2$ !

\begin{table*}[h]
\label{tab:one}
{
\footnotesize 
\centerline{\bf Table 1: Characteristics of different EAS-arrays}
\vspace{0.2cm}
\begin{center}
{\begin{tabular}{|c|c|c|c|c|}
\hline
  Experiment &   Altitude (m) & e.m. Sensitive Area & Instrumented Area & Coverage \\
 & & (m$^2$) & (m$^2$) & \\
\hline
 LHAASO & 4410 & 5.2$\times$10$^3$ & 1.3$\times$10$^6$ & 4$\times$10$^{-3}$ \\
 \hline
 TIBET AS$\gamma$ & 4300 & 380 & 3.7$\times$10$^4$ & 10$^{-2}$ \\
 \hline
IceTop & 2835 & 4.2$\times$10$^2$ & 10$^6$ & 4$\times$10$^{-4}$ \\
\hline
ARGO-YBJ & 4300 & 6700 & 11,000 & 0.93 (central carpet)\\
\hline
KASCADE & 110 & 5$\times$10$^2$ & 4$\times$10$^4$ & 1.2$\times$10$^{-2}$ \\
\hline
KASCADE-Grande & 110 & 370 & 5$\times$10$^5$ & 7$\times$10$^{-4}$ \\
\hline
CASA-MIA & 1450 & 1.6$\times$10$^3$ & 2.3$\times$10$^5$ & 7$\times$10$^{-3}$ \\
\hline
\hline
 & & $\mu$ Sensitive Area & Instrumented Area & Coverage \\
  & & (m$^2$) & (m$^2$) & \\
\hline
LHAASO & 4410 & 4.2$\times$10$^4$ & 10$^6$ & 4.4$\times$10$^{-2}$ \\
 \hline
 TIBET AS$\gamma$ & 4300 & 4.5$\times$10$^3$ & 3.7$\times$10$^4$ & 1.2$\times$10$^{-1}$\\
 \hline
 KASCADE & 110 & 6$\times$10$^2$ & 4$\times$10$^4$ & 1.5$\times$10$^{-2}$ \\
 \hline
 CASA-MIA & 1450 & 2.5$\times$10$^3$ & 2.3$\times$10$^5$ & 1.1$\times$10$^{-2}$ \\
\hline
\end{tabular} }
\end{center}
}
\end{table*}

%
\begin{figure}
  \centerline{\includegraphics[width=0.8\textwidth]{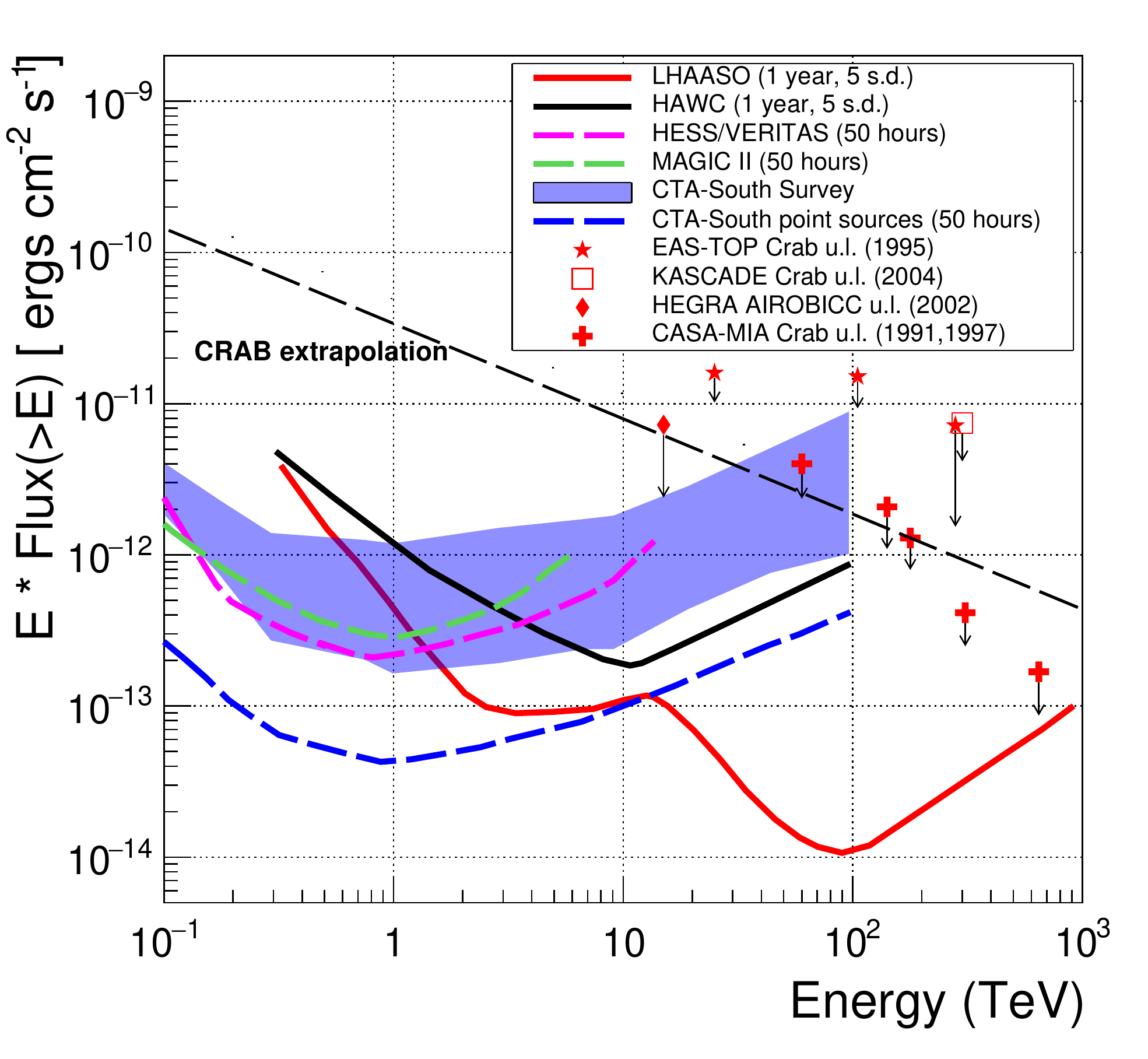} }
\caption{Integral sensitivity of LHAASO as a function of the energy compared to HAWC, HESS, MAGIC II and CTA-South sensitivities. The CTA-South sensitivity to sky survey \cite{cta-survey,hiscore} is also shown.    
Upper limits to ultra high-energy gamma-ray emission set by different experiments in the Northern hemisphere are also reported \cite{eastop97,kascade04,hegra02,casa-mia97}.} 
\label{fig:lhaaso_sens}
\end{figure}
%

Fig. \ref{fig:lhaaso_sens} shows the LHAASO sensitivity in detecting a point--like gamma-ray source, compared to that of other experiments. 
The LHAASO sensitivity shows a structure with two minima, reflecting the fact that the observation and identification of photon showers in different energy ranges is carried out by different detectors: the water Cherenkov array (WCDA) in the range $\sim$0.3 -- 10~TeV and the KM2A array above 10 TeV. 
To compare the sensitivities of the experiments reported in Fig. \ref{fig:lhaaso_sens}, it is important to note that, following
the standard convention, the EAS arrays sensitivity is given for a time of one year, while the sensitivity of Cherenkov telescopes is calculated for 50 hours.
As can be seen in the figure, at energies above few TeV, the LHAASO one-year integral sensitivity is better than the sensitivity of MAGIC or HESS in 50 hours. Above 10--20 TeV the LHAASO sensitivity becomes better than that of CTA-South. At 100 TeV, it is about 30 times better than CTA.
Since an IACT can only spend up to $\sim$200 hours per year observing a single source due to solar and lunar constraints, this means that even with the maximum exposure, an IACT still wouldn't be able to match LHAASO sensitivity above a few tens TeV.

It is important to remark that the LHAASO sensitivity shown in the figure also represents the sensitivity for the survey of a large fraction of the sky (56$\%$ of the celestial sphere for observations with a maximum zenith angle of 40$^{\circ}$), while the sensitivity of Cherenkov telescopes only concerns the observation in a region of a few degrees radius that often contains only one source.  
Every day LHAASO (located at latitude 29$^{\circ}$ North) can survey  the declination band from -11$^{\circ}$ to +69$^{\circ}$
that includes the galactic plane in the longitude interval from +20$^{\circ}$ to +225$^{\circ}$.
The new EAS array HAWC \cite{hawc12} has a sensitivity $\sim$4 times lower than that expected for LHAASO in the 1--10 TeV region, but more than 100 times lower at 100 TeV.
Concerning Cherenkov telescopes, their limited FoV and duty cycle prevent a survey of large regions of the sky.  
To compare the effective sensitivities in sky survey, one has to take into account the fact that a Cherenkov telescope must scan the whole region under study with different pointings. The number of pointings determines the maximum observation time that can be dedicated to any source. 
If we consider a survey of the Galactic Plane in a galactic longitude interval of $\sim$200$^{\circ}$, a reasonable number of pointing is $\sim$100.  Assuming a total observation time of $\sim$1300 hours/year, a full year dedicated to the survey allows an exposure of $\sim$13 hours for source.
This time is reduced to less than 2 hours for an {\it all sky survey} of $\sim\pi$ sr of solid angle, that requires approximately $\sim$800 pointings.
The reduced observation time causes an increase of the minimum detectable flux, as shown in the blue band of Fig.\ref{fig:lhaaso_sens}, where the lower limit of the band refers to a Galactic plane survey and the upper limit to an {\it all sky survey} of $\pi$ sr \cite{cta-survey,hiscore}.

\begin{table*}[h]
{
\footnotesize 
\centerline{\bf Table 2: Performance comparison between LHAASO-KM2A and CASA-MIA experiments}
\vspace{0.3cm}
\label{tab:eascomparison}
\begin{center}
{\begin{tabular}{|c|c|c|c|c|c|}
\hline
  Experiment & Angular resolution & $\mu$ detector                  & EAS array                               & $\mu$ detector & Background hadron \\
                    &                                &     sensitive area (m$^2$)  & instrumented area (m$^2$)    &  coverage        &  surviving efficiency \\
\hline
 LHAASO-KM2A & 0.3$^{\circ}$ (100 TeV) & 42,000   & 1.3$\times$10$^6$ &  4.4$\times$10$^{-2}$ & $\sim$10$^{-5}$ ($\geq$100 TeV) \\
                           &  0.2$^{\circ}$ (1 PeV)    &               &                                &                                     &    \\
\hline
CASA-MIA & 2$^{\circ}$ (100 TeV)              & 2500 & 230,000 & 1.1$\times$10$^{-2}$ & 10$^{-2}$ (178 TeV) \\
                  & $\sim$0.5$^{\circ}$ (646 TeV) &         &               &                                     & 2$\times$10$^{-4}$ (646 TeV) \\
\hline
\hline
 \end{tabular} }
\end{center}
}
\end{table*}

In Fig.\ref{fig:lhaaso_sens} the upper limits set by different experiments to high energy gamma-ray emission in the Northern hemisphere are reported \cite{eastop97,kascade04,hegra02,casa-mia97}.
In five years of observations, the CASA-MIA experiment sets the lowest upper limits to the flux from the Crab Nebula around and above 100 TeV \cite{casa-mia97}. Beyond 1 PeV, the IceTop/IceCube experiments, located at the South Pole, reports a minimum observable gamma ray flux ranging from $\sim$10$^{-19}$ to 10$^{-17}$ photons s$^{-1}$ cm$^{-2}$ TeV$^{-1}$ (depending on the source declination) for sources on the galactic plane in 5 years of measurements \cite{icecube13}.
Table 2 compares the performance of the LHAASO-KM2A array with the CASA-MIA experiment. 
At 100 TeV, the angular resolution of the LHAASO-KM2A array for gamma rays is $\sim$7 times better than that of CASA-MIA, and the area is $\sim$4 times larger. The efficiency in background rejection is about 2$\times$10$^3$ times better in LHAASO, due to the larger muon detector area.
According to expression $F_{min} \propto \frac{ \sigma}{ Q \times \sqrt{A \times T}}$, the LHAASO sensitivity is $\sim$500 times better than that of CASA-MIA at 100 TeV.
With this sensitivity, LHAASO can perform measurements of the high energy tails of emission spectra for the majority of the known TeV galactic sources visible from its location with unprecedented sensitivity.

LHAASO will enable studies in CR physics and gamma-ray astronomy that are unattainable with the current suite of instruments:
\begin{itemize}
\item[1)] LHAASO will perform an \emph{unbiased sky survey of the Northern sky} with a detection threshold better than 10\% Crab units at sub-TeV/TeV and 100 TeV energies in one year. This unique detector will be capable of continuously surveying the $\gamma$-ray sky for steady and transient sources from a few hundred GeV to the PeV energy domain.
From its location LHAASO will observe at TeV energies and with high sensitivity about 30 of the sources catalogued by Fermi-LAT at lower energy, monitoring the variability of 15 AGNs (mainly blazars) at least.
\item[2)] The sub-TeV/TeV LHAASO sensitivity will allow to observe AGN flares that are unobservable by other instruments, including the so-called TeV orphan flares. 
\item[3)] LHAASO will study in detail the high energy tail of the spectra of most of the $\gamma$-ray sources observed at TeV energies, opening for the first time the 100--1000 TeV range to the direct observations of the high energy cosmic ray sources.
\item[4)] LHAASO will map the Galactic \emph{diffuse gamma-ray emission} above few hundred GeV and thereby measure the CR flux and spectrum throughout the Galaxy with high sensitivity. 
The measurement of the space distribution of diffuse $\gamma$-rays will allow to trace the location of the CR sources and the distribution of interstellar gas.
\item[5)] The high background rejection capability in the 10 -- 100 TeV range will allow LHAASO to measure the \emph{isotropic diffuse flux of ultrahigh energy $\gamma$ radiation} expected from a variety of sources including Dark Matter and the interaction  of 10$^{20}$ eV CRs with the 2.7 K microwave background radiation. 
In addition, LHAASO will be able to achieve a limit below the level of the IceCube diffuse neutrino flux at 10 -- 100 TeV, thus constraining the origin of the IceCube astrophysical neutrinos.
\item[6)] LHAASO will allow the reconstruction of the energy spectra of different CR mass groups in the 10$^{12}$ -- 10$^{17}$ eV with unprecedented statistics and resolution, thus tracing the light and heavy components through the knee of the all-particle spectrum.
\item[7)] LHAASO will allow the measurement, for the first time, of the CR anisotropy across the knee separately for light and heavy primary masses.
\item[8)] The different observables (electronic, muonic and Cherenkov components) that will be measured in LHAASO will allow a detailed investigation of the role of the hadronic interaction models, therefore investigating if the EAS development is correctly described by the current simulation codes.
\item[9)] LHAASO will look for signatures of WIMPs as candidate particles for DM with high sensitivity for particles masses above 10 TeV. Moreover, axion-like particle searches are planned, where conversion of gamma-rays to/from axion-like particles can create distinctive features in the spectra of gamma-ray sources and/or increase transparency of the universe by reducing the Extragalactic Background Light (EBL) absorption. 
Testing of Lorentz invariance violation as well as the search for Primordial Black Holes and Q--balls will also be part of the scientific programme of the experiment.
\end{itemize} 

In the next decade CTA-North and LHAASO are expected to be the most sensitive instruments to study Gamma-Ray Astronomy in the Northern hemisphere from about 20 GeV up to PeV.

\subsection{TAIGA-HiSCORE experiment}

The new TAIGA-HiSCORE non-imaging Cherenkov array aims to detect air showers induced by gamma rays above 30 TeV and to study cosmic rays above 100 TeV. 
TAIGA-HiSCORE is made of integrating air Cherenkov detector stations with a wide FoV ($\sim$0.6 sr), placed at a distance of about 100 m to cover a final area of $\sim$5 km$^2$.

%
\begin{figure}
\centerline{\includegraphics[width=0.8\textwidth,clip]{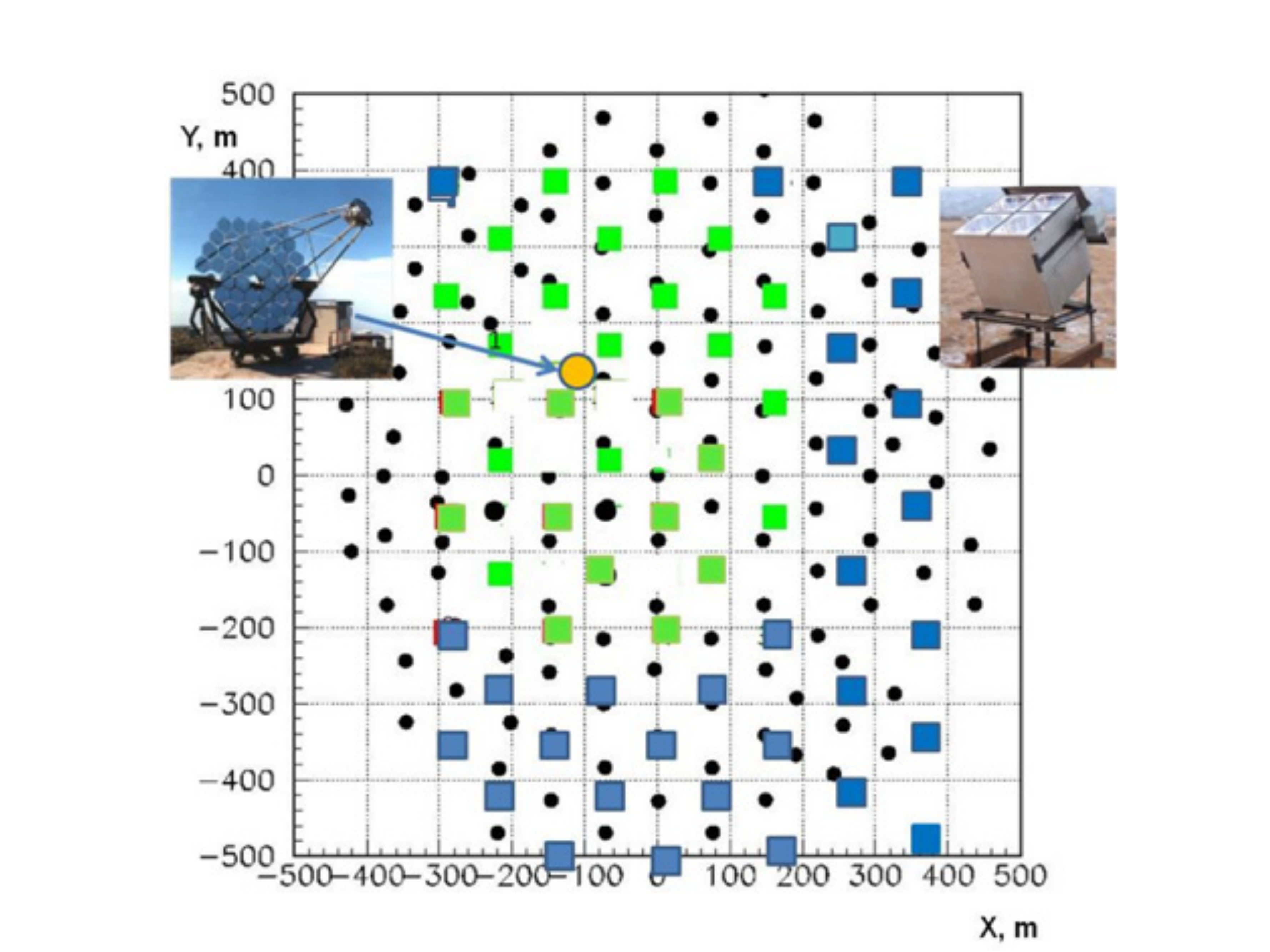} }
\caption{The TAIGA-prototype-2017. Green squares: optical stations of TAIGA-HiSCORE installed in 2014. Blue squares: optical stations to be installed in 2017. Yellow circle: position of the first TAIGA-IACT. Small black circles: optical detectors of Tunka-133.} 
\label{fig:hiscore}       
\end{figure}
%

The TAIGA-HiSCORE array is part of the gamma-ray observatory TAIGA (Tunka Advanced Instrument for cosmic ray physics and Gamma Astronomy). TAIGA is currently under construction in the Tunka valley, about 50 km from Lake Baikal in Siberia, Russia \cite{taiga16}. The key advantage of the TAIGA will be the hybrid detection of EAS Cherenkov radiation by the wide-angle detector stations of the TAIGA-HiSCORE array and by the Imaging Air Cherenkov Telescopes of the TAIGA-IACT array. TAIGA comprises also the Tunka-133 array and will furthermore host up a net of surface and underground stations for measuring the muon component of air showers.
The principle of the TAIGA-HiSCORE detector is the following: the detector stations measure the light amplitudes and full time development of the air shower ligth front up to distances of several hundred meters from the shower core.

Currently TAIGA-HiSCORE array is composed of 28 detector stations distributed in a regular grid over a surface area of 0.25 km$^2$ with an inter-station spacing of about 106 m (prototype array, see Fig. \ref{fig:hiscore}).
Each optical station contains four large area photomultipliers with 20 or 25 cm diameter. Each PMT has a light collector Winston cone with 0.4 m diameter and 30$^{\circ}$ viewing angle (FoV of $\sim$0.6 sr). Plexiglass is used on top to protect the PMTs against dust and humidity. A total station light collection area is 0.5 m$^2$ \cite{hiscore}.

Before the winter season 2017--2018 the TAIGA configuration will include 60 wide angle stations arranged over an area of 0.6 km$^2$, and one single IACT. The expected integral sensitivity for 200 hours of a source observation (about 2 seasons of operation) in the range 30--200 TeV is about 10--12 erg cm$^2$ sec$^{-1}$ \cite{taiga17}.

\section{What's Next ?}
All the experiments mentioned in the previous sections are located in the Northern hemisphere.
The construction of a new wide FoV detector at sufficiently Southern latitude to continuously monitor the Galactic Center and the Inner Galaxy should be a high priority. 
But a new wide FoV detector to be a 'finder' telescope for the future CTA-South experiment needs an energy threshold of $\sim$100 GeV, to be able to detect extragalactic transients (AGNs, GRBs), and a sensitivity at the few percent Crab flux level. 
A 100 GeV energy threshold will also allow to fill the gap between direct and ground-based $\gamma$-ray observations.

Is this possible ?
The main parameters to push down the sensitivity to gamma-ray sources are \cite{discia-icrc17}: (1) the energy threshold; (2) the angular resolution; (3) the gamma/hadron relative trigger efficiency; (4) the effective area for photon detection; (5) the $\gamma$/hadron relative identification efficiency.

The key to lower the energy threshold is to locate the detector at extreme altitude (about 5000 m asl for a threshold in the 100 GeV range). But the energy threshold, as well as the angular resolution, depends also on the coverage (the ratio between the detection area and the instrumented one) and on the granularity of the read-out. The ARGO-YBJ experiment, combining the full coverage approach ($\sim$92\% coverage) at high altitude with a high granularity of the read-out (about 15,000 strips 7$\times$62 cm$^2$ wide), sampled 100 GeV $\gamma$-induced showers with an efficiency af about 70\%. 
The median energy of the first multiplicity bin for photons with a Crab-like energy spectrum was 360 GeV \cite{argo-crab}.
An array of water Cherenkov tanks will hardly be able to lower this energy threshold. Therefore, a full coverage approach based on the Resistive Plate Chambers (RPC) technology continues to be one of the most interesting solution for a new wide FoV in the South.

\subsection{The strong case for a wide FoV telescope in the Southern hemisphere}

The observation of the diffuse emission by AGILE \cite{agile-diffuse} and Fermi \cite{fermi-diffuse} shows that the distribution of CRs above the GeV energy range is smoothly distributed throughout the disc of the Galaxy but the intensity is higher in the Inner Galaxy and falls off towards the outer disc. 
Accordingly, the majority of $\gamma$-ray sources discovered in the last years above 100 GeV have been detected by the Cherenkov telescopes of the H.E.S.S. experiment located in Namibia.
A continuous monitoring of the Inner Galaxy is expected to allow the detection of a large number of new sources.

In addition to the reasons mentioned in Sections 1 and 2, the opportunity to observe the Inner Galaxy offers a lot of additional motivations to push the construction of a wide FoV in the Southern hemisphere.

As mentioned, H.E.S.S. recently claimed the possible detection of a PeVatrons in the Galactic Center, most likely related to a supermassive black hole \cite{hess-pev}, thus opening new perspectives for the observations of CR sources other than SNRs. In fact, the diffuse emission around J1745-290 (positionally compatible with SgrA$^*$) extends up to $\sim$50 TeV, thus suggesting acceleration of CR protons up to the PeV energy range.
Combined with the Fermi-LAT observations of a harder spectrum in the Galactic center region \cite{gaggero17} and the discovery of the so-called Fermi bubbles, we expect that a lot of high-energy particle accelerators are at work in this region that can be studied by a future wide FoV telescope.

Some authors found that the bulk of the Galactic Ridge emission can be naturally explained by the interaction of the diffuse steady-state Galactic CR sea with the gas present in the central molecular zone. Although they confirm the presence of a residual radial-dependent emission associated with a central source, the relevance of the large-scale diffuse component prevents claim solid evidence of a PeVatrons in the Galactic Center to explain the H.E.S.S. data \cite{gaggero17}.
A wide FoV detector may confirm this scenario observing the emission from larger region centered on the Galactic Center.
In addition, the capability of an EAS array to measure the energy spectrum of primary protons up to the PeV energy domain could be crucial to clarify the origin of this emission.

Another interesting opportunity for a wide FoV telescope located in the Southern hemisphere is to investigate the so-called \emph{'IceCube spectral anomaly'}. Recently it was suggested that different IceCube datasets are not consistent with the same power law spectrum of cosmic neutrinos, thus suggesting that they are observing a multicomponent spectrum. 
The passing-muon data (from the Northern sky) agree with isotropy and E$^{-2}$ distribution up to very high energies.
The HESE events (mostly from the Southern sky and shower-like) suggest a softer distribution at low energies. This North-South asymmetry is the IceCube spectral anomaly. It strengthens the likelihood that a Galactic neutrino component exists, mainly observable from the Southern hemisphere.  This Galactic component of neutrinos implies the existence of a Galactic component of very high energy gamma rays \cite{palladino16}.  
A measurement of the $\gamma$-ray flux from the Southern sky in the same energy domain of neutrinos (above 30 TeV) could clarify the origin of the emission. 

Recently, some authors emphasized the existence of an extended 'hot' region of the Southern gamma sky where the cumulative sources contribution dominates over the diffuse component \cite{pagliaroli17}. This region is located in the Inner Galactic Plane
and could be also an important source of HE neutrinos.
Interestingly, this region approximately coincides with the portion of the galactic plane from which a $\sim$2$\sigma$ excess of showers is observed in the HESE IceCube data sample.
If photons are not absorbed, we expect that the high energy neutrino sky is strongly correlated with the high energy gamma sky. A wide FoV detector will allow to study the gamma/neutrino correlation and will provide us a handle to perform a detailed multi-messenger study of the Galactic Plane.

With EAS array some aspects of fundamental physics can be also studied, in particular the existence of multi-TeV Dark Matter particles. The assumption of a single astrophysical power-law flux to explain the IceCube 6-year HESE events yields an anomalous large spectral index ($\gamma^{6yr}$=2.92). Adopting a spectral index in the range [2.0, 2.2], which is compatible with the one deduced by the analysis performed on the 6-year up-going muon neutrinos data, the latest IceCube data show an up to 2.6$\sigma$ excess in the number of events in the energy range 40--200 TeV. Such an excess can be interpreted as a decaying Dark Matter signal \cite{chianese17}. DM particles with masses in the 100 TeV energy range can be investigated only by large wide FoV detectors \cite{lhaaso-dm}.

Finally, a wide FoV detector is an ideal instrument to deepen the observations of the anisotropy in the CR spatial distribution carried out in the Southern hemisphere by IceCube/IceTop experiments \cite{iceaniso} and to measure energy spectrum and elemental composition of the CR primary flux.

\subsection{The LATTES proposal}

Different groups are proposing different ideas for a new wide FoV in the South \cite{alto,alpaca,mostafa}.

An interesting proposal is to built a hybrid detector to exploit the characteristics of two detectors, RPCs and Water Cherenkov Detectors (WCD).
The LATTES project will consist of one layer of RPCs on top of WCD of small dimensions \cite{lattes}.
%
\begin{figure}
\centerline{\includegraphics[width=0.8\textwidth,clip]{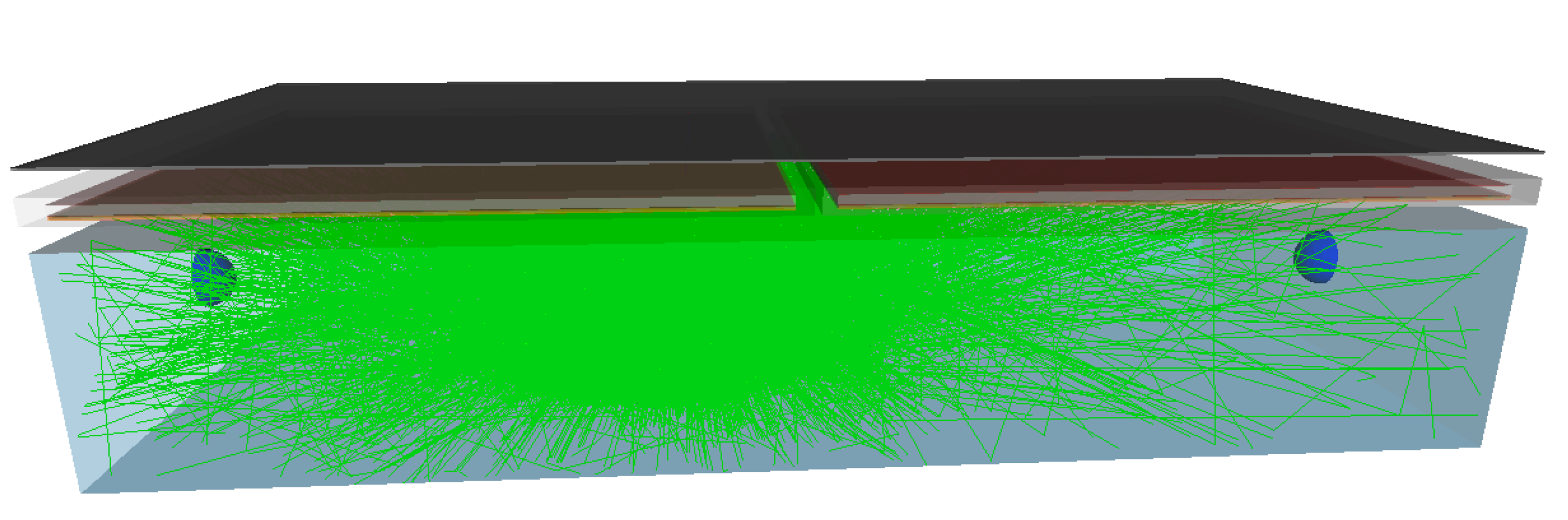} }
\caption{Basic LATTES detector station, with one WCD covered with RPCs and a thin slab of lead \cite{lattes}. 
The green lines show the tracks of the Cherenkov photons inside the water.} 
\label{fig:lattes-basic_station}       
\end{figure}
%

The basic element, shown in Fig. \ref{fig:lattes-basic_station}, is composed of one WCD, with a rectangular horizontal surface of 
3 m$\times$1.5 m and a depth of 0.5 m, covered by two RPCs, each with a surface of  (1.5$\times$1.5) m$^2$, with a lead layer on top (5.6 mm) to exploit the secondary photons conversion.
The proposed RPCs are based on glass, different from the bakelite RPCs operated for more than 10 years at 4300 m asl by the ARGO-YBJ collaboration \cite{discia-rev} and extensively used at LHC.

The proposed RPCs are designed to work at low gas flux, (1--4) cc/min, at harsh outdoor environment, and demanding very low maintenance services. Their intrinsic time resolution was measured to be better than 1 ns.
The WCD read-out will be provided by two 15 cm PMTs at both ends of the smallest vertical face of the tank.
The read-out of the RPCs will be provided by 16 charge collecting pads.
This hybrid detector is expected to improve the trigger selection at low energies and the rejection of the background of charged nuclei.
The shower energy will be reconstructed from the total signal, defined as the sum of the number of photoelectrons in all WCD stations.
The proposed experiment will consist of an array of 60$\times$30 stations, covering an effective area of about 10,000 m$^2$ located at 5200 m asl.
The different detectors will be separated by a small distance (roughly 0.5 m) to allow access to PMTs and RPCs. 
An angular resolution better than 2$^{\circ}$ is expected in the 100 GeV range.
A 1-year sensitivity at level of 15\% of the Crab Nebula flux is expected in the 100 -- 400 GeV energy range.

\section{Conclusions}

Open problems in cosmic ray physics push the construction of new generation EAS arrays to study, in the 10$^{11}$ -- 10$^{18}$ eV energy range, at the same time photon- and charged-induced events.

Multi-messenger astronomy, the exploration of the Universe through combining information from different cosmic messengers: electromagnetic radiation, gravitational waves, neutrinos and cosmic rays, will greatly benefit from an air shower wide FoV detectors.
As shown by the analysis of the recent Gravitational Wave event GW170817, only a wide-angle telescope is able to survey simultaneously all the large sky regions identified by LIGO and VIRGO, looking for a possible correlated $\gamma$-ray emission.

LHAASO is the most ambitious project for a new generation multi-component wide FoV experiment in the Northern hemisphere.
In the next decade CTA-North and LHAASO are expected to be the most sensitive instruments to study gamma-ray astronomy in the Northern hemisphere from about 20 GeV up to PeV.

A new EAS array to study the 100 GeV $\gamma$-sky with high sensitivity and to monitor the Galactic Center should be a high priority.
Extreme altitude ($\sim$5000 m asl), high coverage coupled to a high granularity of the read-out are the key to improve the angular resolution and the sensitivity to gamma-ray sources.

The ARGO-YBJ Collaboration demonstrated that bakelite RPCs can be safely operated at extreme altitudes for many years. 
The benefits in the use of RPCs in ARGO-YBJ were: (1) high efficiency detection of low energy showers (energy threshold $\sim$300 GeV) by means of the dense sampling of the central carpet; (2) unprecedented wide energy range investigated by means of the digital/charge read-outs ($\sim$300 GeV $\to$ 10 PeV); (3) good angular resolution and unprecedented details in the core region by means of the high granularity of the read-outs.

The LATTES project is a hybrid detector to exploit both the ARGO-YBJ full coverage approach with RPCs and the HAWC/LHAASO technique to reject the background of charged cosmic rays with a water Cherenkov detector.
Such an experiment, if located in the Southern hemisphere at 5200 m asl with an instrumented area of 10,000 m$^2$, can reach a sensitivity of order of 10\% of the Crab Nebula flux in one year in the 100 GeV energy range, thus monitoring the Galactic Center and  complementing the coming Cherenkov Telescope Array.

\end{document}